\documentclass[12pt]{article}
\usepackage{graphicx}
\usepackage{amsmath}
\usepackage{amssymb}
\usepackage{epsfig}

\def\pbnr{}
\def\speaker{Sasa Prelovsek}
\def\title{Lattice QCD review of charmonium and open-charm spectroscopy}
\def\affiliation{Jozef Stefan Institute,  Ljubljana, Slovenia\\
University of Ljubljana, Slovenia}
\def\support{ sasa.prelovsek@ijs.si}
\newcommand\pubnumber{\pbnr}
\newcommand\pubdate{\today}
\def\Title#1{\begin{center} {\Large #1 } \end{center}}
\def\Author#1{\begin{center}{ \sc #1} \end{center}}

\newcommand{\OnBehalf}[1]{\sbox0{#1}\ifdim\wd0=0pt
        {}% if #1 is empty
	\else
	{\\on behalf of #1}% if #1 is not empty
	\fi}
\newcommand{\SupportedBy}[1]{\sbox0{#1}\ifdim\wd0=0pt
        {}% if #1 is empty
	\else
	{\footnote{#1}}% if #1 is not empty
	\fi}
\def\Address#1{\begin{center}{ \it #1} \end{center}}

\newcommand\pubblock{\includegraphics[width=5cm]{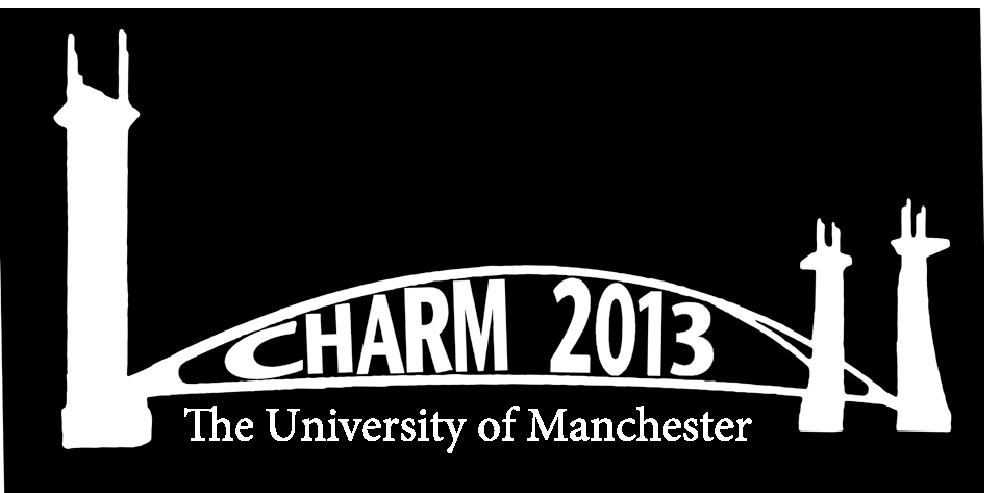}\hfill{\begin{tabular}{l} \pubnumber\\
         \pubdate  \end{tabular}}}
\newenvironment{Abstract}{\begin{quotation}  }{\end{quotation}}
\newenvironment{Presented}{\begin{quotation} \begin{center} 
             PRESENTED AT\end{center}\bigskip 
      \begin{center}\begin{large}}{\end{large}\end{center} \end{quotation}}
\def\Acknowledgements{\bigskip  \bigskip \begin{center} \begin{large}
             \bf ACKNOWLEDGEMENTS \end{large}\end{center}}
\def\venue{The 6$^{th}$ International Workshop on Charm Physics\\
(CHARM 2013)\\
Manchester, UK,  31 August -- 4 September, 2013}
\def\beq{\begin{equation}}
\def\eeq#1{\label{#1}\end{equation}}
\def\eeqn{\end{equation}}

\def\beqa{\begin{eqnarray}}
\def\eeqa#1{\label{#1}\end{eqnarray}}
\def\eeqan{\end{eqnarray}}

\let\bar=\overbar

\def\Dslash{\not{\hbox{\kern-4pt $D$}}}
\def\dslash{\not{\hbox{\kern-2pt $\del$}}}

\def\msb{{\bar{\ssstyle M \kern -1pt S}}}

\begin{document}
\begin{titlepage}
\pubblock

\vfill
\Title{\title}
\vfill
\Author{\speaker\SupportedBy{\support} }
\Address{\affiliation}
\vfill
\begin{Abstract}
%%%%%%%%%%%%%%%%%%%%%%%%%%%%%%%%%%%%%%%%%%%%%%%%%%%%%%%%%%%%%%%%%%%%%%%%%%%
% YOUR ABSTRACT GOES HERE
%%%%%%%%%%%%%%%%%%%%%%%%%%%%%%%%%%%%%%%%%%%%%%%%%%%%%%%%%%%%%%%%%%%%%%%%%%% 
Lattice QCD results on the spectroscopy of charmonium(like) states and open charm mesons are reviewed. Near-threshold states $X(3872)$ and  $D_{s0}^*(2317)$ were treated rigorously for the first time and the searches for $Z_c^+(3900)$, $X(4140)$ and $cc\bar u\bar d$ were carried out. The  first simulations of the resonances with charm quarks have been performed, including the determination of their strong decay widths. Spectroscopy of highly excited  charmed, charmonium and hybrid states has been calculated. 
\end{Abstract}
\vfill
\begin{Presented}
\venue
\end{Presented}
\vfill
\end{titlepage}
\def\thefootnote{\fnsymbol{footnote}}
\setcounter{footnote}{0}
%

%%%%%%%%%%%%%%%%%%%%%%%%%%%%%%%%%%%%%%%%%%%%%%%%%%%%%%%%%%%%%%%%%%%%%%%%%%%
%  WHAT FOLLOWS IS YOUR TEXT
%%%%%%%%%%%%%%%%%%%%%%%%%%%%%%%%%%%%%%%%%%%%%%%%%%%%%%%%%%%%%%%%%%%%%%%%%%%
\section{Introduction and motivation} 

 The masses of states well below the open charm threshold, which starts at $m_D+m_\pi$ for $D$ mesons and at $2m_D$ for charmonium, have been reliably and precisely calculated  on the lattice for several years. With exception of \cite{Bali:2011rd}, the states near threshold or above threshold have been treated using the so-called ``single-meson'' approach until 2012, which essentially means that effect of the threshold and strong decays of these states were ignored.  
The impressive spectrum of high lying quark-antiquark and hybrid multiplets was calculated during the past two years using this less-complete treatment. In addition, the first exploratory lattice simulations of the near-threshold states and resonances based on the rigorous treatment were done during the last year. These are the main focus of this talk. 

Proper treatment of near-threshold states in lattice QCD is particularly important since most exotic experimental candidates lie near open-charm thresholds. Examples include $X(3872)$ within  $1~$MeV of $D^0D^{0*}$ threshold. The structure called  $Z_c^+(3900)$ was discovered in $J/\psi \pi^+$ invariant mass in 2013 by BESII \cite{Ablikim:2013mio} and confirmed by Belle and CLEOc \cite{Liu:2013dau,Xiao:2013iha}.  It   has quark flavor structure $\bar cc\bar du$ and also lies close to $DD^*$ threshold. Other charged charmonium-like  states $Z^+(4430)$ and more recent $Z_c^+(4020)$ \cite{Ablikim:2013wzq}, $Z_c^+(4025)$ \cite{Ablikim:2013emm} have been observed only in one experiment.  They also have flavor $\bar cc\bar du$ and lie near  $D^*D_1$ and $D^*D^*$ thresholds. The only lattice study aimed at $Z_c(3900)$ will be reviewed, while $Z^+(4430)$ was addressed only in \cite{Meng:2009qt}. 

Masses are extracted from the energy levels $E_n$, and these are obtained from the correlation functions $C_{ij}(t)=\langle O_i(t)O_j(0)\rangle=\sum_n \langle O_i|n\rangle \langle O_j|n\rangle^* e^{-E_n t}$ evaluated on the lattice, where interpolating fields $O$ create and destroy the physical state with given quantum number $J^{PC}$. The simulations of charmonium(like) states discussed below neglect the charm annihilation Wick contractions in view of OZI suppression. For hadrons with charm quarks it is common to compare $m-m_{reference}$ between lattice and experiment, where the leading discretization errors related to $m_c$ cancel. 

\begin{figure}[htb]
\centering
$\quad$
\includegraphics*[width=0.85\textwidth,clip]{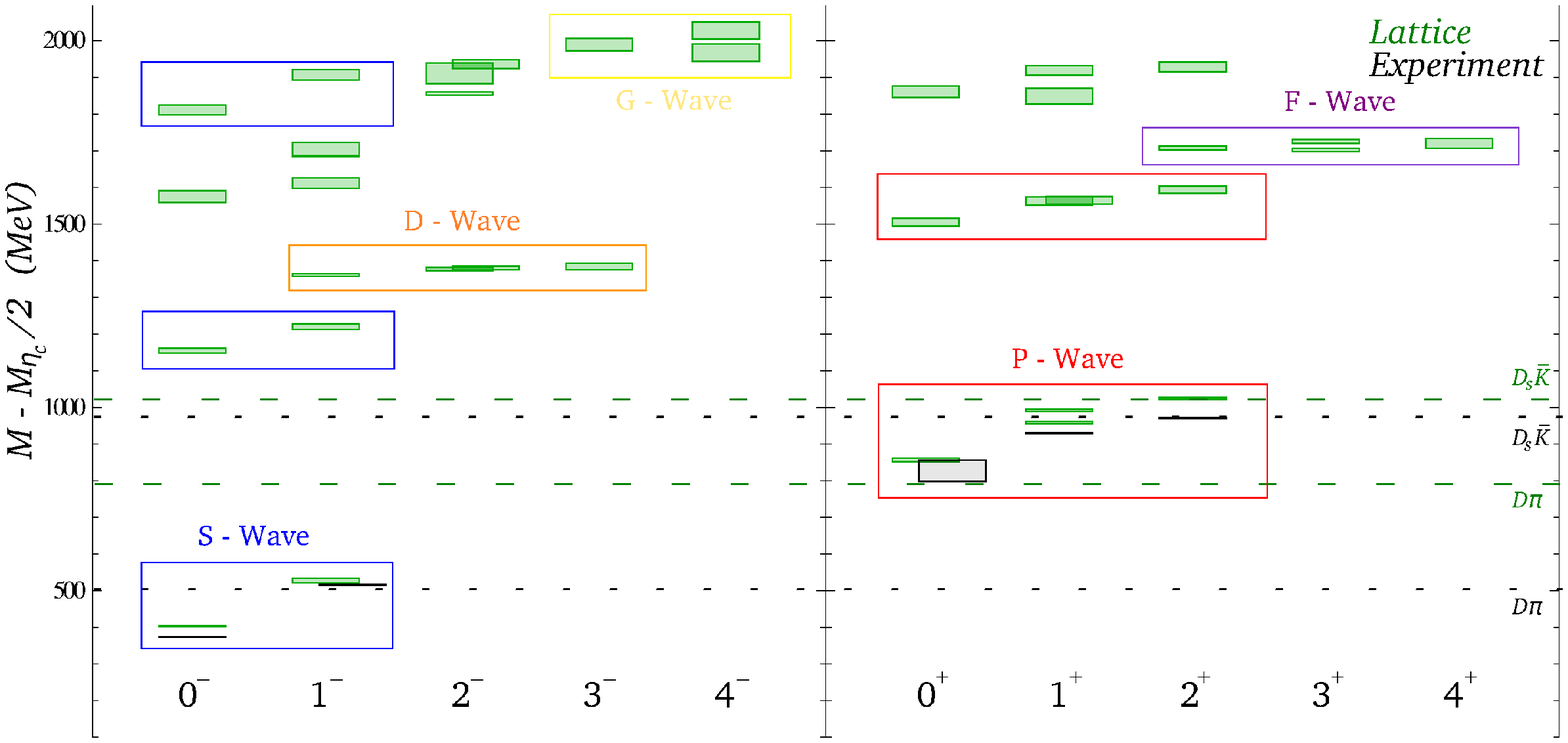}

$\quad$
\includegraphics*[width=0.85\textwidth,clip]{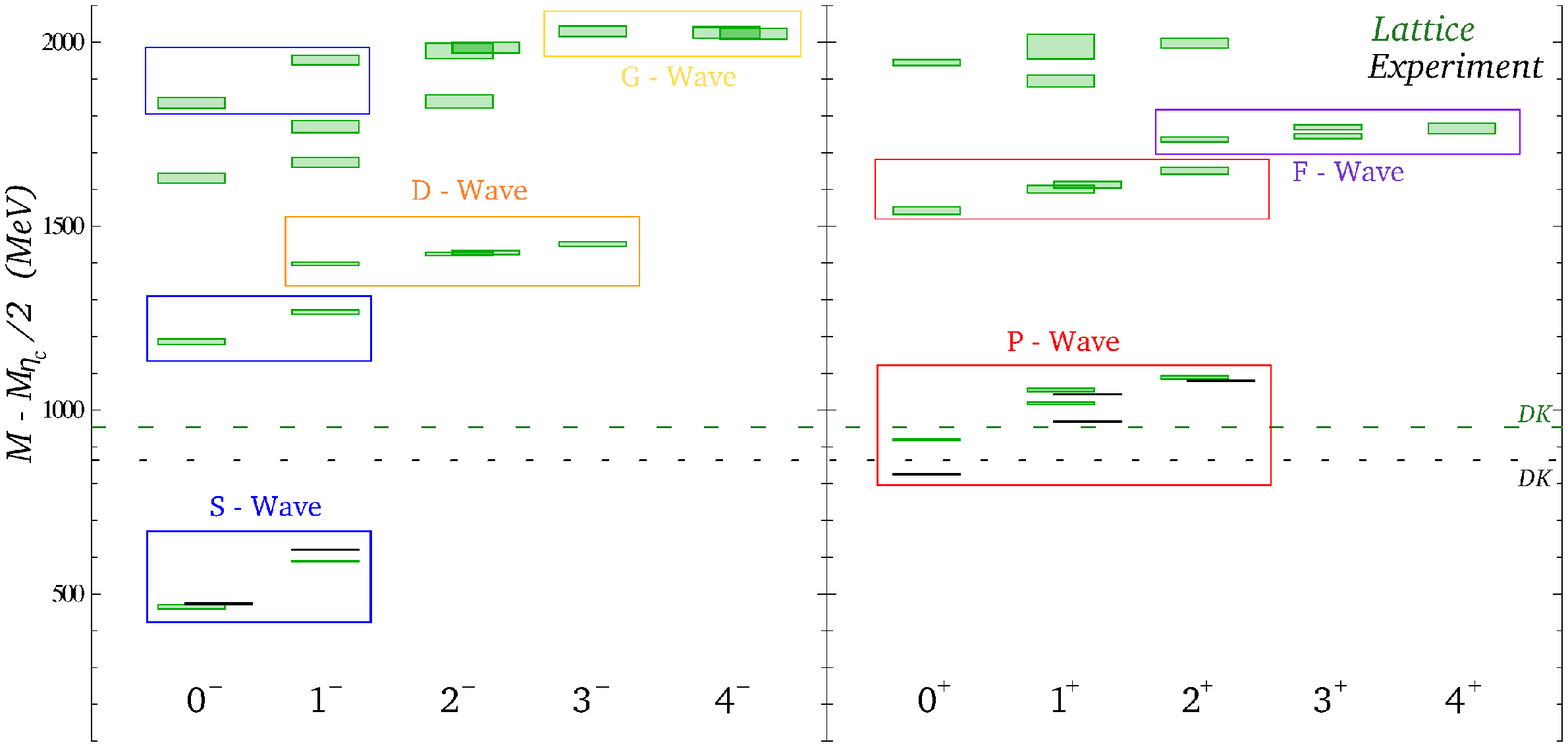}
\caption{\label{fig:hsc_D} The $D$ (above) and $D_s$ (below) spectra and interpretation from HSC  at $m_\pi\!\simeq\! 400~$MeV \cite{Moir:2013ub} (plot from G. Moir).   }
\end{figure} 

\begin{figure}[t!]
\centering
\includegraphics*[width=0.80\textwidth,clip]{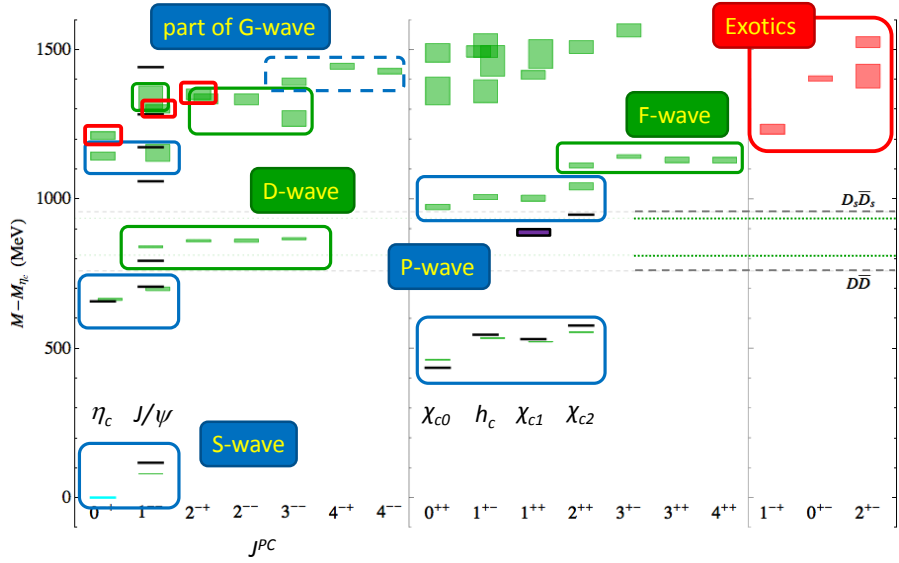} 
\caption{\label{fig:hsc_cc} The $c\bar c$ spectrum and interpretation from the HSC at $m_\pi\! \approx\! 400~$MeV \cite{Liu:2012ze} (plot from talk of C. Thomas at this meeting).}
\end{figure} 

\begin{figure}[b!]
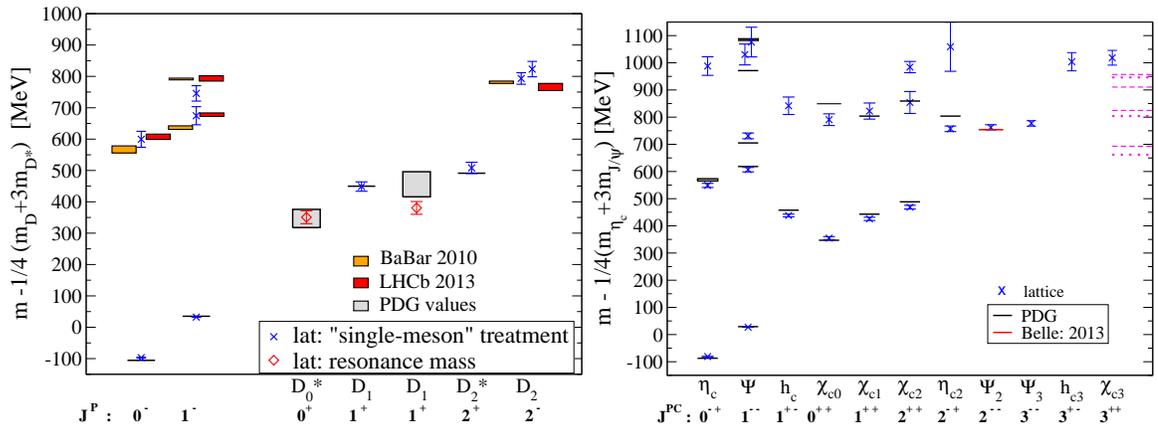

\centering
\includegraphics*[width=0.50\textwidth,clip]{figs/lattice_exp_D_2.eps}
\includegraphics*[width=0.49\textwidth,clip]{figs/lattice_exp_cc.eps}
\caption{\label{fig:Dnaive}  The $D$ and $\bar cc$ spectrum extracted in \cite{Mohler:2012na} at $m_\pi\!\simeq\! 266~$MeV.   }
\end{figure} 

\section{States well below open charm thresholds}

The recipe for extracting a mass of a state well below the strong decay threshold is straightforward. The mass  $m\!=\!\sqrt{E^2-P^2}$ in a simulation with $P\!=\!0$ is equal to the energy level $E$. The resulting mass  is extrapolated to $a\!\to\! 0$ and $L\!\to\! \infty$, while the quark masses are extrapolated or interpolated to the physical value. Particular care needs to be taken about the discretization errors related to the $c$ quarks and it has been verified that independent methods lead to compatible results in the $a\!\to\! 0$ limit. 

 The precise masses for low-lying states were obtained recently after extrapolations by HPQCD \cite{Dowdall:2012ab,Donald:2012ga},  Briceno {\it et al.} \cite{Briceno:2012wt}, FNAL/MILC \cite{DeTar:2012xk} and $\chi$QCD \cite{chiqcd:tmp}, while related studies are   in \cite{Mohler:2012gn,Becirevic:2012dc,Namekawa:2011wt,Kalinowski:2013wsa}. The resulting masses or mass splittings are typically within  $10~$MeV from the experimental value.

\section{``Single-meson'' treatment of excited states}

The ``single-meson'' treatment of states near or above threshold refers to (i) using only quark-antiquark interpolating fields $O\sim \bar qq$ for mesons, (ii) assuming that all energy levels correspond to ``one-particle'' states and (iii) that the mass of the state equals the measured energy level  $m\!=\!E$. These are strong assumptions for the resonances, which are not asymptotic states.  The approach also ignores the effect of the threshold on near-threshold states.

The most extensive $D$, $D_s$ \cite{Moir:2013ub} and $\bar cc$ \cite{Liu:2012ze} spectrum, shown in Figs. \ref{fig:hsc_D} and \ref{fig:hsc_cc}, was extracted  by the Hadron Spectrum Collaboration (HSC) on $N_f\!=\!2\!+\!1$ anisotropic configurations with $m_\pi\simeq 400~$MeV and two different $L\simeq 2.9~$fm,~$1.9~$fm.  The continuum $J^{PC}$ was reliably identified using advanced spin-identification method. An impressive number of excited states was extracted in each channel with a good accuracy.  States  are identified with members of $\bar qq$ multiplets $1S$, $2S$, $3S$, $1P$, $2P$, $1D$ and $1F$ based on overlaps $\langle O_i|n\rangle$, where interpolators are chosen to resemble multiplet members.  There are several remaining states, which are not part of  $\bar qq$ multiplets and these states show particularly strong overlap with interpolators $O\simeq \bar q F_{\mu\nu}q$   containing the gluon field strength tensor. These states are identified as hybrids and also turn out to form several full multiplets. Some of the identified charmonium hybrids have manifestly exotic $J^{PC}$, while the others have conventional values. The open-charm meson hybrids have non-exotic $J^P$.

The $D$ meson spectrum in Fig. \ref{fig:Dnaive} from a simulation with $m_\pi\simeq 266~$MeV \cite{Mohler:2012na} shows reasonable agreement with 2010 Babar  \cite{delAmoSanchez:2010vq} and 2013 LHCb \cite{Aaij:2013sza} excited states, which are shown side by side. The blue crosses result from the ``single-meson'' treatment, while the red diamonds are resonances masses from the rigorous treatment. The splitting between $1S$ and $2S$ multiplets of the order $\simeq 650~$MeV is in agreement  experiments.  The new $\bar cc$ state $2^{--}$ \cite{Bhardwaj:2013rmw} agrees with the prediction in Fig. \ref{fig:Dnaive}.   

Other recent related results can be found in \cite{Bali:2012ua,Bali:2011dc,Mohler:2011ke,Kalinowski:2013wsa}.

\section{Rigorous treatment of resonances}

\begin{figure}[htb]
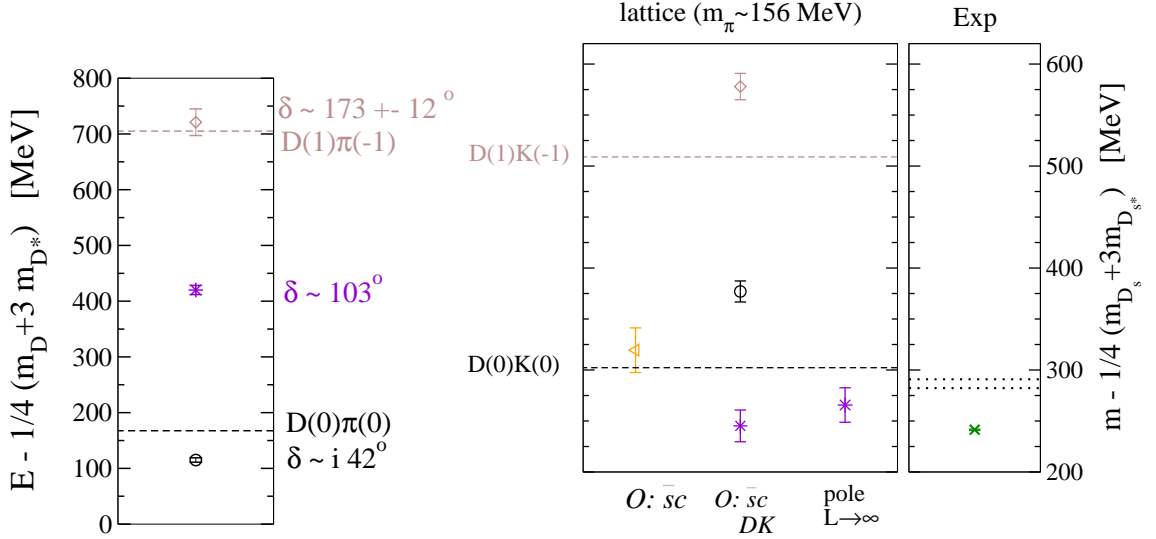

\centering
\includegraphics*[width=0.39\textwidth,clip]{figs/delMeV_Dscalar.eps}
\includegraphics*[width=0.59\textwidth,clip]{figs/Ds.eps}
\caption{\label{fig:Ds}  Left: the energy levels and $D\pi$ phase shifts in $D^{*}_0(2400)$ channel with $I\!=\!1/2$ and $J^P\!=\!0^+$ \cite{Mohler:2012na}. Right: the levels in $D_{s0}^*(2317)$ channel with $I\!=\!0$ and $J^P\!=\!0^+$ \cite{Mohler:2013rwa}.     }
\end{figure}

\underline{$\mathbf{D_{0}^*(2400)}$ {\bf and} $\mathbf{D_{1}(2430)}$}: Let me illustrate the basics of the rigorous treatment for the example of 
 $D^{*}_0(2400)$ scalar resonance, which appears in $D\pi$ scattering in s-wave.  All states with given quantum number $J^P\!=\!0^+$ in principle appear as energy levels:
 \begin{itemize} 
 \item
 Important examples are two-particle states $D(p)\pi(-p)$, where periodic boundary conditions require  $p\!=\! n\tfrac{2\pi}{L}$  in absence of the interaction. The corresponding energy $E\!=\!\sqrt{m_1^2+p^2}+\sqrt{m_2^2+p^2}$ of non-interacting two-particle states will be marked by horizontal lines in the figures. 
 The meson-meson interpolating fields need to be incorporated to make sure these levels are seen in practice. 
 \item 
 The levels that appear in addition to the expected two-particle states are related to the bound states or resonances. 
 \end{itemize}
The first and the third levels in Fig. \ref{fig:Ds} correspond to the interacting  $D\pi$ two-particle states, while the second level is related to $D^{*}_0(2400)$ resonance.   
  Each energy level $E$ renders an elastic phase shift $\delta(E)$ at that energy via the rigorous L\"uscher relation \cite{Mohler:2012na}. The three energy levels  give the three phase shift points around the $D^{*}_0(2400)$ resonance region. The Breit-Wigner type fit through these three points leads to the resonance mass in Fig. \ref{fig:Dnaive} very close the the experiment, and the resonance width reasonably close as well \cite{Mohler:2012na}. The mass and width of the broad $D_1(2430)$ resonance in $D^*\pi$ scattering is also close to experiment (see Fig. \ref{fig:Dnaive}), but the analysis in this channel relies on few assumptions due to two nearby   $D_1(2430)$ and $D_1(2420)$  \cite{Mohler:2012na}.   
 
\vspace{0.2cm}

\underline{$\mathbf{\chi_{c0}^\prime }$}: An indication for a charmonium $J^{PC}\!=\!0^{++}$ resonance with $m_{\chi_{c0}^\prime}= 3932 \pm 25~$MeV and $\Gamma(\chi_{c0}^\prime \to \bar DD)= 36\pm17~$MeV and additional enhancement of $\sigma(D\bar D)$ near threshold in a simulation \cite{prelovsek:tmp} prompts experiments to look for such structures in  $D\bar D$ invariant mass (see \cite{Chen:2012wy,Guo:2012tv} for similar suggestions).%meissner,chen   

\begin{figure}[htb]
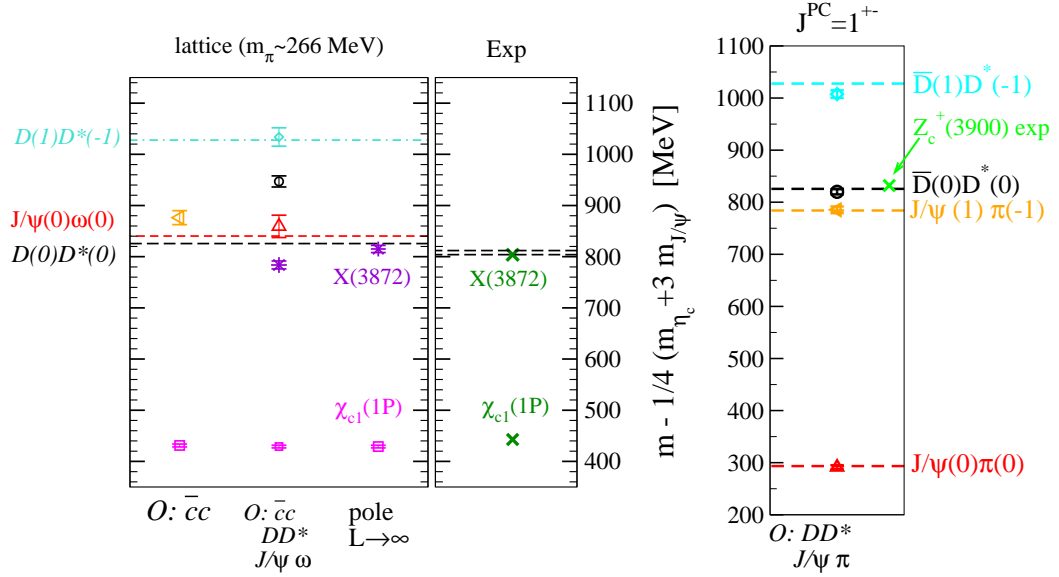

\centering
\includegraphics*[width=0.6\textwidth,clip]{figs/delMeV_T1pp_I0_charm13.eps}
\includegraphics*[width=0.3\textwidth,clip]{figs/Zc.eps}
\caption{\label{fig:X} Left: the energy levels in $X(3872)$ channel with $I\!=\!0$ and $J^{PC}\!=\!1^{++}$ \cite{Prelovsek:2013cra}. Right: the levels in $Z_c(3900)$ channel with $I\!=\!1$ and $J^{PC}\!=\!1^{+-}$ \cite{Prelovsek:2013xba}.   }
\end{figure} 

\section{Rigorous treatment of near-threshold states}

\underline{$\mathbf{D_{s0}(2317)}$}: Fig. \ref{fig:Ds} shows lowest three energy levels from the  first simulation of $D_{s0}^*(2317)$ that takes into account the effect of the $DK$ threshold by explicitly incorporating $DK$ interpolating fields \cite{Mohler:2013rwa}. The position of the $DK$ threshold is almost physical in this $N_f\!=\!2+1$ simulation with nearly physical  $m_\pi\simeq 156~$MeV.
 The second and third levels are related to the interacting $DK$ states, while the first level is related to 
 $D_{s0}^*(2317)$ and it is below threshold as in experiment. 
The lowest two levels lead to the  scattering length  $a_0\!=\!-1.33\pm 0.20~$fm and small effective range $r_0$ for $DK$ scattering  in s-wave. The negative $a_0$ is an indication for a the presence of $D_{s0}^*(2317)$  below threshold \cite{Sasaki:2006jn}.   The effective range expansion leads to the position of the pole at $L\!\to\! \infty$, rendering  $m_{D_{s0}^*}-\tfrac{1}{4}(m_{D_s}+3m_{D_s^*})=266\pm 16$~MeV  close to the experimental value $241.5\pm 0.8~$MeV. 

The ``single-meson'' treatment using just $\bar sc$ interpolators gives  only one level in Fig. \ref{fig:Ds} (orange triangles) with a misleading energy and one can not reliably establish whether this level corresponds to $D_{s0}^*(2317)$ or $D(0)K(0)$.  

\vspace{0.2cm}

 \underline{$\mathbf{X(3872)}$}: A candidate for the charmonium(like) state $X(3872)$ is found $11\pm 7~$MeV below the $D\bar D^*$ threshold for $J^{PC}\!=\!1^{++}$, $I\!=\!0$, $N_f\!=\!2$ and $m_\pi\!\simeq\! 266~$MeV  \cite{Prelovsek:2013cra}. This is the first lattice simulation that establishes a candidate for $X(3872)$ (violet stars in Fig. \ref{fig:X}) in addition to $\chi_{c1}$ (squares)  and the nearby scattering states $D\bar D^*$ (circles and diamonds) and $J/\psi\,\omega$ (triangles). The established $X(3872)$ has a large overlap with $\bar cc$ as well as $D\bar D^*$ interpolating fields \cite{Prelovsek:2013cra}. The large and negative $a_0\!=\!-1.7\pm 0.4~$fm for $D\bar D^*$ scattering  is an indication for a shallow bound state $X(3872)$ \cite{Sasaki:2006jn}. 

The ``single-meson'' treatment using just $\bar cc$ interpolators  renders only one level in Fig. \ref{fig:X} (orange triangles) near $DD^*$ threshold just like in  previous simulations.  One can not reliably establish whether this level is related to $X(3872)$ or $D(0)\bar D^*(0)$.

In the  $I\!=\!1$ channel, only the $D\bar D^*$ and $J/\psi\,\rho$ scattering states are found, and no candidate for $X(3872)$ \cite{Prelovsek:2013cra}. This is in agreement  with a popular interpretation that $X(3872)$ is dominantly $I\!=\!0$, while its small $I\!=\!1$ component arises solely from the isospin breaking and is therefore absent in the simulation with $m_u\!=\!m_d$.  

\section{Search for exotic states}

 \underline{$\mathbf{Z_c(3900)}$}: Three experiments recently reported a discovery of a manifestly exotic  $Z_c^+(3900)$ in the decay to $J/\psi\;\pi^+$ \cite{Prelovsek:2013xba}. It has $C=-$, while $J$ and $P$ are experimentally unknown. The first search for this interesting state on the lattice focused on a channel with $J^{PC}\!=\!1^{+-}$ and $I\!=\!1$. Fig. \ref{fig:X} indicates that only the expected two-particle states $D\bar D^*$ and $J/\psi\; \pi$ are found. No additional level was found that could be related to   $Z_c^+(3900)$.  The possible reasons for not finding $Z_c^+$ may be that its $J^{PC}$ are not $1^{+-}$ or that the employed interpolating fields are not diverse enough. Simulation with additional types of interpolators will be needed to reach a more definite conclusion. 

\vspace{0.2cm}

\begin{figure}[htb]
\centering
\includegraphics*[width=0.45\textwidth,clip]{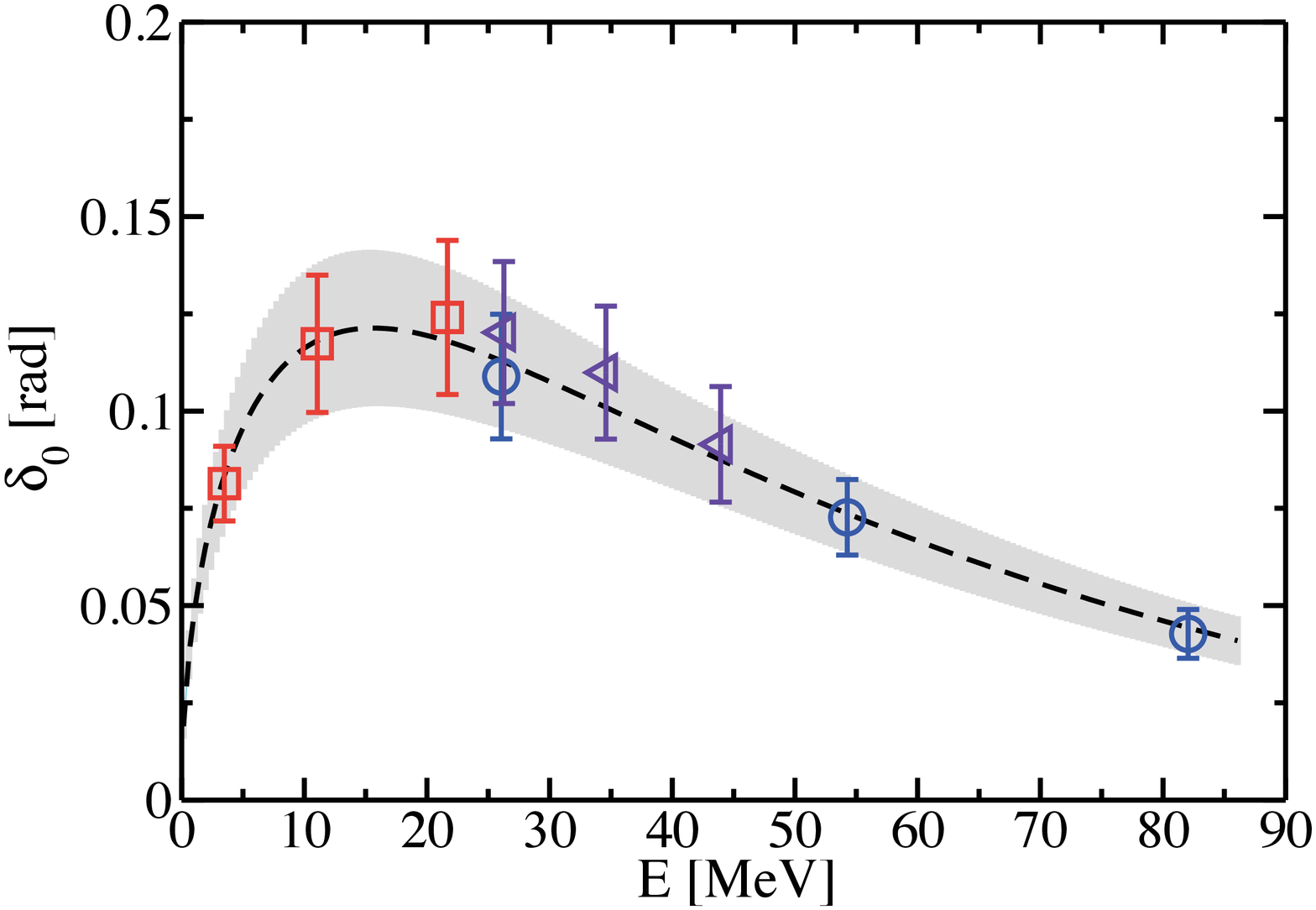}
\includegraphics*[width=0.45\textwidth,clip]{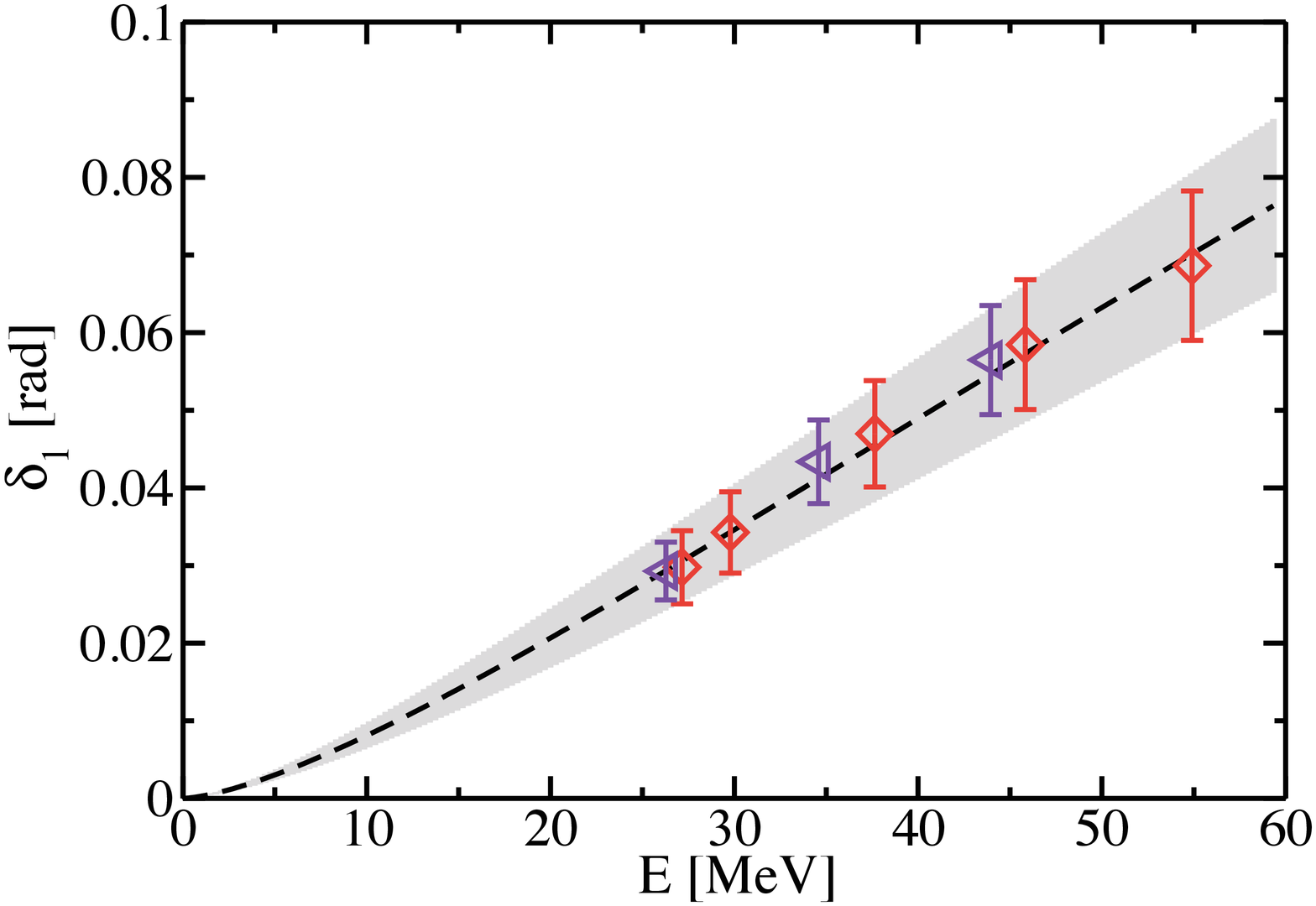}
\caption{\label{fig:sasaki}  The phase shift for s-wave and p-wave scattering of $J/\psi\;\phi$ from \cite{Ozaki:2012ce}.   }
\end{figure}

\underline{$\mathbf{X(4140)}$}: A structure called $X(4140)$ was found in $J/\psi\;\phi$ invariant mass by  CDF \cite{Aaltonen:2009tz} and more recently by CMS  and D0 \cite{Chatrchyan:2013dma,Abazov:2013xda}. The s-wave and p-wave $J/\psi\;\phi$ scattering phase shift in Fig. \ref{fig:sasaki} was extracted  in \cite{Ozaki:2012ce} ignoring $\bar ss$ annihilation contribution, and no resonant structure was found.  

\vspace{0.2cm}

\begin{figure}[htb]
\centering
\includegraphics*[width=0.52\textwidth,clip]{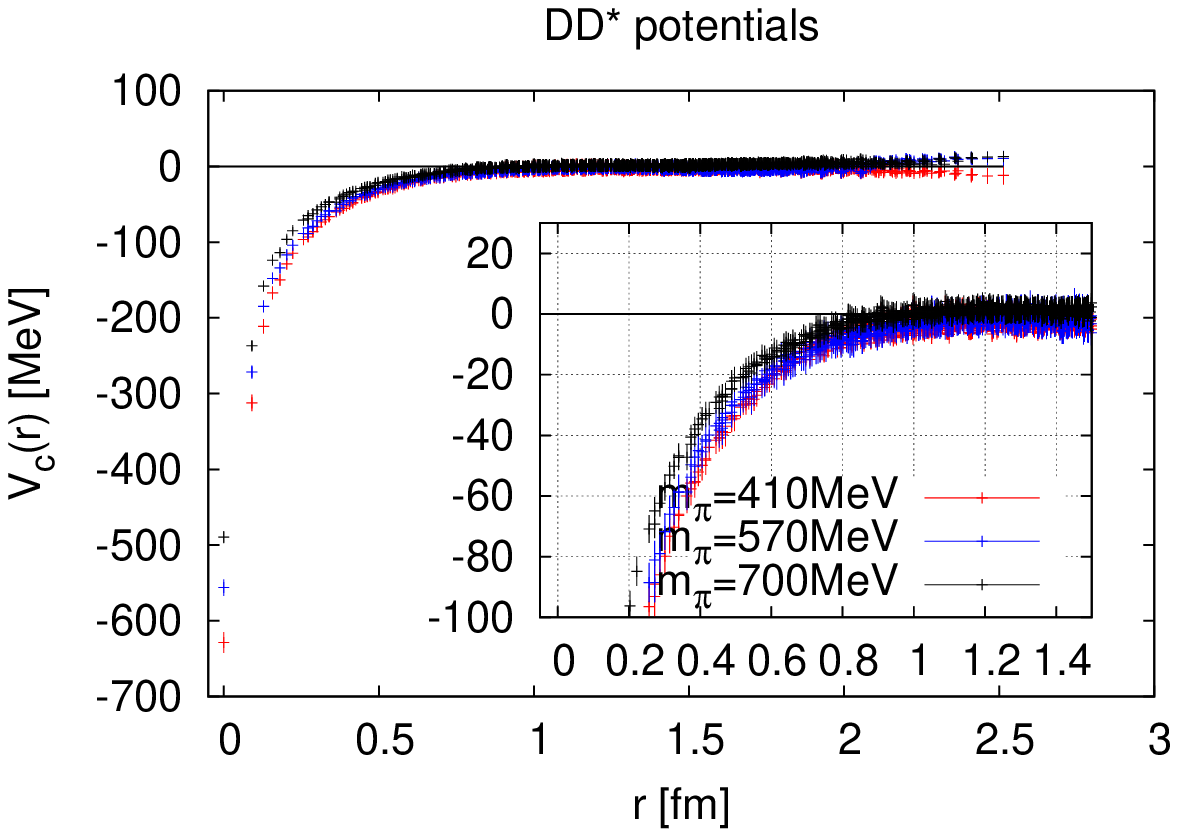}
\includegraphics*[width=0.47\textwidth,clip]{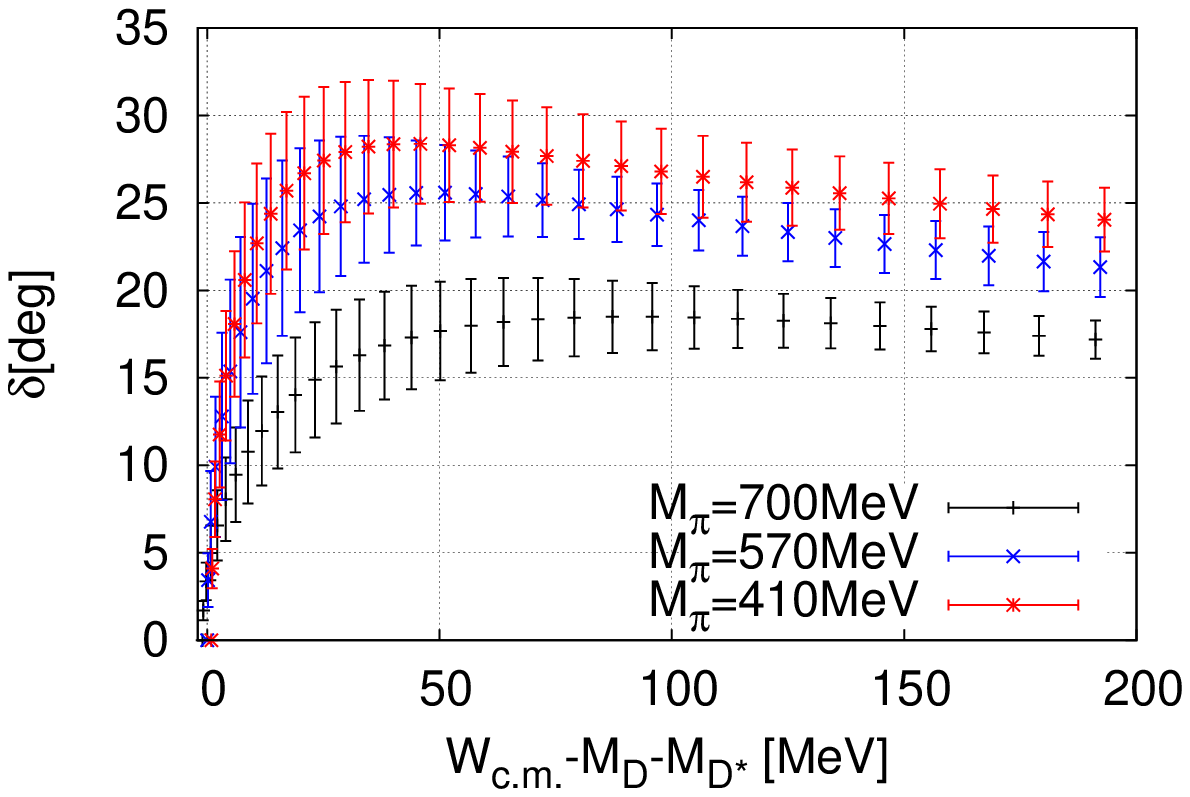}
\caption{\label{fig:halqcd} The $DD^*$ potential and phase shift in double-charm channel \cite{halqcd:tmp}.   }
\end{figure}

\underline{$\mathbf{cc\bar d\bar u}$ {\bf tetraquark}}:
The potential $V(r)$ between $D\!=\!\bar uc$ and $D^*\!=\!\bar dc$ at distance $r$ was extracted using the HALQCD time-dependent method \cite{HALQCD:2012aa}. The resulting potential in  Fig. \ref{fig:halqcd} is attractive, but the corresponding $DD^*$ scattering phase shift does not start at $\delta(0)=\pi$, which indicates that there is no $cc\bar d\bar u$ tetraquark bound state at the simulated $m_\pi$. 

\section{Related topics}

\underline{\bf Interactions of  charmed mesons with light pseudoscalars}: The scatt\-er\-ing length  was extracted in five non-resonant channels for four different $m_\pi$ in a simulation \cite{Liu:2012zya}. The simultaneous fit to these channels renders the low-energy constants of the SU(3) Unitarized ChPT. These give indirect prediction for the resonant $DK$ scattering with $I\!=\!0$ and $J^+\!=\!0^+$, where the pole is found at $m_{D_{s0}^*}=2315{+18\atop -28}~$MeV close to the experimental $D_{s0}(2317)$ \cite{Liu:2012zya}.  

\vspace{0.2cm}

\begin{figure}[htb]
\centering
\includegraphics*[width=0.6\textwidth,clip]{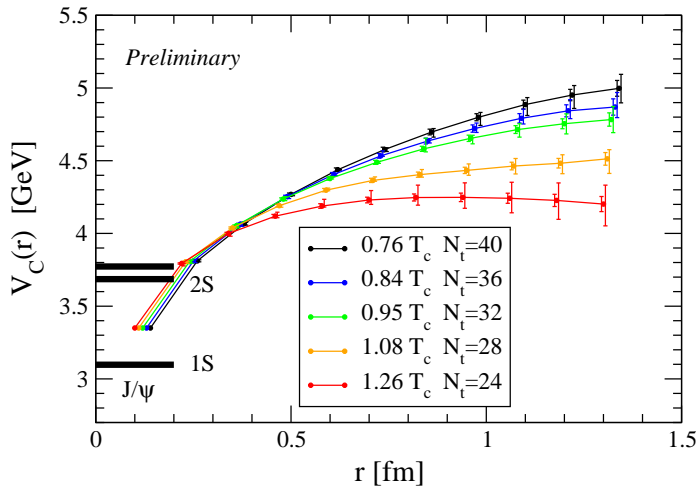}
\caption{\label{fig:nonzeroT} The spin-independent part of the $\bar cc$ potential in $N_f\!=\!2+1$ simulation  \cite{Evans:2013zca}.}
\end{figure}

\underline{\bf Charmonium potential at finite temperature}: 
The $V_C(r)$ part of the potential $V(r)=V_C(r)+\vec s_1\cdot \vec s_2~V_S(r)$ between $\bar c$ and $c$ at distance $r$ was extracted using the HALQCD time-dependent method at non-zero temperatures  in $N_f\!=\!2+1$ simulation \cite{Evans:2013zca}. The resulting potential in Fig. \ref{fig:nonzeroT} is temperature dependent and becomes flatter at large $r$ as the temperature increases.  

\section{Conclusions}

The states well below open charm decay threshold are reliably and precisely determined from lattice QCD. During the past two years, the impressive spectrum of $D$, $D_s$ and $\bar cc$ quark-antiquark and hybrid multiplets was calculated within the simplified single-meson approach. Several  near-threshold states and resonances were treated rigorously for the first time: candidates for $D_{s0}^*(2317)$, $X(3872)$, $D_0(2400)$ and $D_1(2430)$ were established, while $Z_c(3900)$, $X(4140)$ and bound $cc\bar d\bar u$ tetraquarks were not found (yet). Precision simulations of these channels in the future would be valuable.

\Acknowledgements
I would like to thank my collaborators C.B. Lang, L. Leskovec, D. Mohler and R. M. Woloshyn.  Thanks to everyone who sent me the material for this review. This work is supported by ARRS project number N1-0020 and FWF project number I1313-N27. 

\bibliographystyle{h-physrev4}
\bibliography{Lgt}

\end{document}